\documentclass[aps,prd,showpacs,preprintnumbers]{revtex4}
\usepackage{graphics}
\def \beq{\begin{equation}}
\def \eeq{\end{equation}}
\def \beqa{\begin{eqnarray}}
\def \eeqa{\end{eqnarray}}
\newcommand{\cO}{{\cal O}}

\def \etal{{\sl et al.\/}}
\def \jhep{{\sl J.\ H.\ E.\ P.\/}}
\def \np{{\sl Nucl.\ Phys.\/}}
\def \pl{{\sl Phys.\ Lett.\/}}
\def \pr{{\sl Phys.\ Rev.\/}}
\def \prl{{\sl Phys.\ Rev.\ Lett.\/}}
\def \ie{{\sl i.\ e.\/}}

\begin{document}
 
\title{Here's looking at you, fireball}
\author{Sourendu \surname{Gupta}}
\email{sgupta@theory.tifr.res.in}
\affiliation{Department of Theoretical Physics, Tata Institute of Fundamental
         Research,\\ Homi Bhabha Road, Mumbai 400005, India.}

\begin{abstract}

Plasmas screen electromagnetic waves of frequency less than the plasma
frequency, $\omega_p$, with a skin depth, $\delta$, specified by the
electrical conductivity, $\sigma$. Current estimates of the transport
properties of the QCD plasma indicates that photons with energy less
than 250--500 MeV would be screened with a skin depth of 2--4 fm. In
a hadron gas, a little below $T_c$, screening occurs for much softer
photons, of energy less than approximately 50 MeV, albeit with similar
values of $\delta$. If the QCD plasma is indeed strongly interacting,
then it can be proven by merely looking at the brightness of a fireball
from different angles.

\end{abstract}
\pacs{\hfill
TIFR/TH/04-30, hep-lat/0411355}
\maketitle

With clear signatures of dense matter formation in heavy-ion collisions
at the RHIC \cite{rhic}, attention must shift to the characterization of
this matter. One fact is self-evident--- this matter consists of mobile
electrically charged constituents, and is therefore an electromagnetic
plasma. All such plasmas screen photons. The photon skin depth, $\delta$,
is related to the electrical conductivity of the plasma, $\sigma$. Photons
of frequency $\omega$ are screened for $\omega<\omega_p$, where $\omega_p$
is the plasma frequency. There is evidence that the plasma formed at
RHIC is strongly interacting. If so, the skin depth should be fairly
small, and soft photon emission should be approximately a surface effect,
leading to strong variations of intensity when the fireball is observed
from different angles.

{\bf The hot phase}:
Photon spectra were first computed in weak-coupling theory almost
two decades ago, and the state of the art has improved continuously
\cite{weak}. There was even a computation of $\sigma/T$ for the
plasma some years back: the leading behaviour is $1/g^4\log(1/g)$
and the coefficient of this term was computed \cite{amy}. However,
these computations, especially for soft photons, may not be easily
adapted to RHIC conditions. The reason is that a careful analysis of
the QCD scale in finite temperature lattice computations reveals that
$T_c/\Lambda_{\overline{\rm MS}} =0.49\pm0.02$ \cite{precise}, where $T_c$
is the cross-over temperature in QCD. As a result, even at $T=10T_c$,
the QCD running coupling $\alpha_S =0.14$, when evaluated at the scale
$2\pi T$. This implies that $g>1$.

The weak-coupling computations depend on the smallness of $g$ to separate
the length scales $1/T\ll 1/gT\ll 1/g^2T\cdots$ for small $g$. When $g>1$
this scale separation does not work, and may even give qualitatively
incorrect results. Direct tests for the separation of these scales
through lattice computations show a complete breakdown of the argument
since magnetic screening length scales (expected to be order $1/g^2T$)
are shorter than electric screening length scales (expected to be of
order $1/gT$) in the QCD plasma for $T\approx2$--3$T_c$ \cite{saumen}.
In other words, a controlled weak-coupling computation of the equation
of state and transport coefficients such as the viscosity and electrical
conductivity is not possible in the vicinity of $T_c$.

Nevertheless, agreement between phenomenological analyses of data
and lattice QCD results is emerging: it has become clear that the
high-temperature phase of QCD ($T>T_c$, where $T_c$ is the crossover
temperature for QCD matter) is strongly dissipative. There is strong
evidence for short thermalization times, which requires rapid dissipation
\cite{kolb}. A recent fit to data was able to extract the shear viscosity,
$\eta$, from RHIC data on particle spectra and flow \cite{teaney}. This
gives $\eta/S\approx0.14$, where $S$ is the entropy density in the
plasma. A recent lattice computation in quenched QCD implies that
$\eta/S\approx0.2$ \cite{sgupta}. It has been conjectured that the
universal strong coupling limit of this quantity is $\eta/S\ge1/4\pi$
\cite{son}, implying that QCD is close to this limit.

\begin{figure}
\begin{center}
   \scalebox{0.4}{\includegraphics{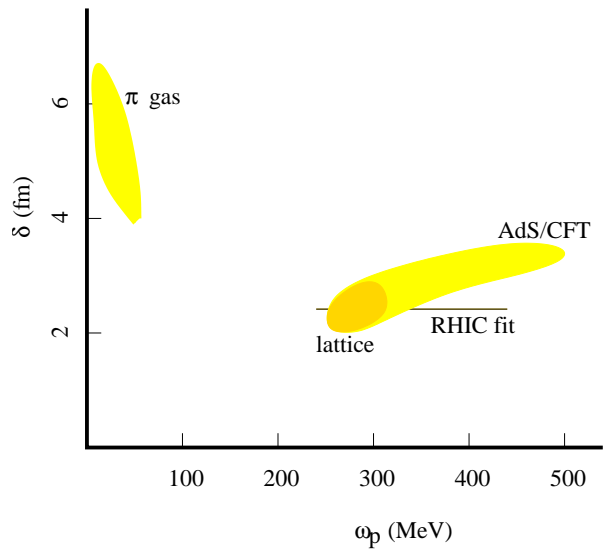}}
\end{center}
\caption{Regions of the parameter space, \ie, the plasma frequency, $\omega_p$,
   and the skin depth, $\delta$, likely in different phases of the plasma.
   The indicated numbers are discussed in the text--- the darker region is
   the most likely region of initial parameters in the hot phase of QCD. The
   system is likely to move towards the AdS/CFT limit closer to $T_c$. If
   the EOS near $T_c$ is very soft, then the fireball lingers there, and
   photons of energy up to 500 MeV may be screened.}
\label{fg.limits}\end{figure}

In the plasma phase of QCD, the electric charge carriers are quarks, and
the momentum carriers are dominantly gluons. As a result, the electrical
conductivity, $\sigma\propto\tau_q$ whereas the shear viscosity,
$\eta\propto\tau_g$, where $\tau_{q,g}$ are effective mean-free times for
quarks and gluons \cite{note1}. However, the two interaction strengths
are related by the gauge symmetry of the theory.  As a result, the
(dimensional) ratio $\sigma/\eta$, should be completely specified by
the temperature and chemical potentials in the QCD plasma.  To put this
another way, two different estimates of $\eta$ in a certain ratio to
each other would give rise to two estimates of $\sigma$ in the same ratio.

An estimate of $\sigma$ can be directly used to compute the skin depth
for a soft photon by plugging the constitutive relation, $j=\sigma E$,
\ie, Ohm's law, into Maxwell's equations \cite{jackson} to give
\beq
   \delta=\sqrt{\frac2{\omega_p\sigma}}.
\label{skinhot}\eeq
A lattice computation estimated $\sigma\approx T/7$ and
$\omega_p\approx0.94T$ \cite{sgupta}. This implies that the skin depth
is rather small, with $\delta\approx2$ fm at $\omega_p$.  One can scale
this for the value of $\eta$ estimated in \cite{teaney}, to obtain
$\delta\approx2.5$ fm. Finally, if the bound in \cite{son} is saturated
by the QCD plasma, then one should find a somewhat larger skin depth---
$\delta\approx3.5$ fm. In this last case, the plasma frequency would be
$\omega_p\approx500$ MeV.

\begin{figure}
\begin{center}
   \scalebox{0.6}{\includegraphics{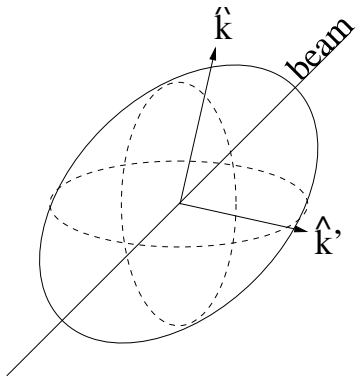}}
\end{center}
\caption{Photon intensities for a fixed impact parameter and different
   angles $\hat{\bf k}$ and $\hat{\bf k}'$. The ellipsoid is a representation
   of the fireball at one instant of time. For realistic skin depths the
   brightness of the source would depend on the direction from which it is
   viewed. For moderate skin depths, the brightness is roughly proportional
   to the cross sectional areas, which are shown as the dashed ellipses.}
\label{fg.angles}\end{figure}

{\bf The cold phase}:
The cold phase of this hadronic matter exists in a narrow band
of temperature--- between $T_c$ and the freeze out temperature
$T_f$.  Current estimates of $T_f$ are bounded by $m_\pi<T_f<2m_\pi$
\cite{cleymans}, as a result of which thermal pair production of charged
pions is extremely unlikely. In fact,
\beq
   \frac1{\exp(E_\pi/T)-1}={\rm e}^{-m_\pi/T}{\rm e}^{-m_\pi v^2/2T}
          + \cO\left({\rm e}^{-2m_\pi/T}\right),
\label{correct}\eeq
and hence the hadron gas may be considered to be a classical
non-relativistic plasma within an accuracy of about 30\%. The next
lightest particle is the kaon. Since $m_K/m_\pi>2$, this same argument
suffices to show that its contribution to transport coefficients is
negligible at this level of approximation.  Thus, as far as the transport
properties are concerned, the cold phase can be considered approximately
as a classical non-relativistic two component electromagnetic plasma,
with charge carriers $\pi^\pm$ mixed in with the uncharged $\pi^0$.

In a Maxwell-Boltzmann plasma, light is screened in the medium with
a skin depth of the order of the inverse plasma frequency, \ie,
\beq
   \delta=1/\omega_p, \qquad{\rm for}\qquad 
   \omega<\omega_p = \left(\frac{8\pi\alpha n_\pi}{m_\pi}\right)^{1/2},
\label{simple}\eeq
where $\alpha$ is the fine structure constant and $n_\pi$ is the number
density of each species of charge pions. We have taken into account the
fact that there are two oppositely charged species of pions. There are
numerically small corrections to this estimate, due to the change in the
wavenumber from medium effects, leading to the estimate
\beq
   \delta = \frac1{\omega_p} \left[\left(\frac{2T}{\pi m_\pi}\right)^{1/2}
      \frac{\omega_p}\omega\right]^{1/3}.
\label{skincold}\eeq
for $\omega<\omega_p$ \cite{ichimaru}.  For pions to be the appropriate
degrees of freedom, there can be no more than one pion in every box of
side equal to a pion Compton wavelength. Since $n_\pi$ above counts only
one species of pions, by this argument it should equal 1/3 per pion
Compton wavelength. In this high density limit one finds screening
for photon energies less than 50 MeV with a skin depth of 3.9 fm
or more.  As $n_\pi$ decreases $\omega_p$ decreases as $\sqrt{n_\pi}$
and simultaneously the skin depth increases. Thus, for more realistic
charged pion densities, screening would occur at even smaller energies,
and with larger skin depth. Note that there is no particular relation
between momentum and charge transport time scales for a pion gas. As a
result, the AdS/CFT limit is not an upper bound to $\delta$.

Collective effects may become strong as $n_\pi$ approaches the limit
of one pion per Compton wavelength. If this were to happen at low
temperature then one might expect condensation phenomena of various
kinds which could tend to lower the pion mass and lead to a decrease in
$\omega_p$. However, $T_c$ is not small enough for condensation to occur.
Instead, the limit $n_\pi\to1$ becomes strongly interacting, and possibly
leads to the universal limit of \cite{son}. This scenario preserves
the holy cow of continuity at the QCD crossover, $T_c$.

Finally, we give an estimate of the error due to the neglect of the
heavier mesons. If one takes into account the full hadronic spectrum,
then there is a contribution to $\omega_p^2$ of $4\pi\alpha e_h^2 n_h/m_h$
from each hadron $h$.  Since the density of heavier hadrons is smaller
at any $T\simeq m_\pi$, these give exponentially small corrections
in $m_h/m_\pi$.

{\bf Experimental signatures}:
As we have estimated above, very soft photons, of energy at most
$\omega_p=$250--500 MeV, would be required to see the skin depth of a
plasma. This poses a technical challenge of rejecting all the photons
which are produced by decays after free streaming of hadrons sets in.
This is a problem I leave to people who know the detectors best. The
question I ask here is--- if this separation can be done reliably,
then what could one observe?

Let us define a quantity called the contrast which takes values between
zero and unity---
\beq
   \kappa(\delta) = \left.\frac{N(\hat{\bf k};\delta)-N(\hat{\bf k}';\delta)}
         {N(\hat{\bf k};\delta)+N(\hat{\bf k}';\delta)}\right|_{max},
\label{contrast}\eeq
where $N(\hat{\bf k};\delta)$ is the photon intensity observed in the
direction $\hat{\bf k}$ (specified by the pseudo-rapidity $\eta$ and
azimuthal angle $\phi$) when the skin depth is $\delta$, and the ratio
on the right is maximized by varying over two directions, $\hat{\bf k}$
and $\hat{\bf k}'$, independently within the limits of the experimental
acceptance. The photon intensity entering the definition above can be
computed by the formula---
\beq
   N(\hat{\bf k};\delta) = \int d^4x \frac{d^4N(T)}{dx^4}
          {\rm e}^{-\ell({\bf x}\hat{\bf k})/\delta},
\label{intensity}\eeq
where the integral is over the 4-volume of the fireball, the photon
emission rate depends implicitly on the point of emission, ${\bf x}$,
through the local temperature,  $T({\bf x})$, and the path-length of
the photon through the fireball is a purely geometric quantity which is
completely specified by the point of origin, ${\bf x}$, and the direction
of propagation, $\hat{\bf k}$. At the level of this approximation, all the
interaction effects are taken into $\delta$. It would be appropriate then
to take $d^4N(T)/dx^4$ as that for a blackbody at temperature $T$.

Two limits can be obtained analytically. First, for photon energies large
enough that the electrical conductivity of the plasma can be neglected,
the skin depth can be set to infinity. Then the exponential factor in eq.\
(\ref{intensity}) is unity, and the full volume is observable from any
angle. The contrast vanishes in this limit--- $\kappa(\infty)=0$. This
is exactly what happens for $\omega\gg\omega_p$.

In the opposite limit, where the conductivity is very large, \ie,
in the limit of vanishing $\delta$, the exponential factor becomes a
delta function at the freezeout surface. Since the photon emission rate
is completely specified by the temperature, it can be pulled out of the
integral, which then becomes a product of the surface area and a surface
brightness. The latter factor is a function only of $T$, and hence is
constant over the freezeout surface. As a result, the contrast becomes a
purely geometric quantity, being proportional to the surface area visible
from the detector element in the direction $\hat{\bf k}$. It is clear that
if the surface is a sphere then the contrast vanishes--- the detector
sits macroscopically far away, and therefore, from any angle sees one
hemisphere, \ie, one half of the total surface area. This generalizes. If
the freezeout surface has an inversion symmetry about the center and
is strictly convex, then from any viewpoint one sees half the surface
\cite{note2}. In this limit again, the contrast vanishes--- $\kappa(0)=0$.

The contrast does not vanish in general for intermediate values of
$\delta$.  A simple example is a translucent chapati (thin disk)--- face
on the full volume may be visible, but edge on only a small fraction is
visible. However, even in this case, some extremely symmetric shapes such
as a sphere or an infinitely long tube give vanishing contrast. 

As we have argued above, when the skin depth is finite. freezeout is
no longer confined to a surface. However, the notion of isothermal
surfaces in the fireball (of which the freezeout surface is an example)
still exists and is useful.  We have presented the arguments above for
space-like freezeout isotherms, but a generalization to the time-like
case is straightforward. It also becomes clear from the above discussion
why it is necessary to distinguish photons emitted before decoupling from
photons produced by decays of particles after decoupling. An analogous
problem in astronomy is to separate fluctuations of the primordial
radiation from later events like stars or nearby gas clouds.

Due to the high degree of symmetry, central collisions yield the smallest
contrast.  The contrast can be increased by selecting non-central
collisions. Since there seems to be a range of impact parameters where
the initial energy density changes little, it is possible to improve
the contrast without sacrificing the initial energy density, and hence
the signal which depends on the contrast.  At the RHIC (and the LHC)
the skin depth could be 30--50\% of the fireball dimensions \cite{size}.
As a result, some contrast should be visible.  If it is, then eq.\
(\ref{intensity}) can be used in conjunction with realistic models such
as a full hydro code (preferably with dissipation built in, at least to
first order) to extract the value of $\delta$.

It is interesting to note that the contrast may be boosted by using
asymmetric ion combinations, say S-Au or Pb-U. In this case, the fireball
no longer has inversion symmetry, and one can have $\kappa(0)>0$ (think of
the difference between an egg and an ellipsoid). Since $\kappa(\infty)=0$
independent of the shape of the fireball, the qualitative fact of a
skin-depth smaller than the fireball dimension can be proven (or ruled
out) by using an asymmetric ion combinations. No fit or detailed
modelling is needed.

There is one other point that requires comment. In the discussion of
the contrast, $\kappa$, we have assumed that the freezeout surface has
uniform properties. However, if a jet punches through, then it makes
spurious contributions to the contrast. These events should be removed
from the sample which is analyzed. In the present fairly advanced stage
of jet physics at the RHIC, it should be possible to do this efficiently.

The point of this paper is the following--- in the high-temperature phase
of QCD gauge symmetry relates the reaction rates which are relevant
to different transport coefficients such as the shear viscosity and
the electrical conductivity.  Since there is growing evidence that
the viscosities and other dissipative time scales in QCD matter are
short, this has implications for the electrical conductivity, and
hence observational consequences as a small skin depth for photons of
frequency less than or equal to the plasma frequency. I have suggested
one method for the observation of the skin depth here, through a study
of the contrast. However, the underlying physics is important enough that
one should explore other ways of investigating it even if the contrast is
not easily visible--- varying the ion species used and the beam energies,
or using asymmetric ion combinations, are all ideas worth exploring.



\begin{thebibliography}{99}
\bibitem{rhic}
  See, for example,
  M.\ Gyulassy and L.\ McLerran, nucl-th/0405013;
  R.\ Rapp, nucl-th/0403048.
\bibitem{weak}
  For recent reviews see
  F.\ Gelis, \np, A 715 (2003) 329;
  F.\ Arleo \etal, hep-ph/0311131;
  P.\ Aurenche, hep-ph/0410282.
\bibitem{amy}
   P.\ Arnold, G.\ D.\ Moore, L.\ G.\ Yaffe, \jhep, 0112 (2001) 009.
\bibitem{precise}
   S.\ Gupta, \pr, D 64 (2001) 034507.
\bibitem{saumen}
   S.\ Datta and S.\ Gupta, \np, B534 (1998) 392.
\bibitem{kolb}
  U.\ W.\ Heinz and P.\ F.\ Kolb, hep-ph/0204061.
\bibitem{teaney}
  D.\ Teaney, \pr, C 68 (2003) 034913.
\bibitem{sgupta}
  S.\ Gupta, \pl, B 597 (2004) 57.
\bibitem{son}
  G.\ Policastro, D.\  T.\ Son and A.\ Starinets, \prl, 87 (2001) 081601.
\bibitem{note1}
   Very close to $T_c$, lattice computations imply that the assumption that there
   are gluon-like quasi-particles in the plasma is false \cite{saumen2}. The
   conjecture that the plasma is full of coloured composites \cite{shuryak} would
   give rise to interesting transport phenomena, including quantitative predictions
   of the electrical conductivity, which can then be compared to lattice results.
\bibitem{saumen2}
   S.\ Datta and S.\ Gupta, \pr, D 67 (2003) 054503.
\bibitem{shuryak}
   E.\ Shuryak, {\sl Prog.\ Nucl.\ Part.\ Phys.\/}, 53 (2004) 273.
\bibitem{jackson}
  J.\ D.\ Jackson, {\sl Classical Electrodynamics\/}, (1999), John Wiley
  and Sons, Singapore.
\bibitem{cleymans}
  See, for example, the freezeout curve in J.\ Cleymans \etal, hep-ph/0411187.
\bibitem{ichimaru}
  S.\ Ichimaru, {\sl Basic Principles of Plasma Physics: a Statistical
  Approach\/}, Frontiers in Physics Series (1973) Benjamin/Cummins
  Publishing Company, Reading, Massachusetts, USA.
\bibitem{note2}
   The importance of inversion symmetry was pointed out to me by Nitin Nitsure.
\bibitem{size}
  C.\ Adler \etal, (STAR Collaboration), \prl, 87 (2001) 082301;\\
  S.\ S.\ Adler \etal, (PHENIX Collaboration), \prl, 93 (2004) 152302.
\end{thebibliography}
\end{document}